\documentclass[11pt, a4paper]{amsart}
\usepackage[dvips]{epsfig}
\usepackage{amsmath}
\usepackage{amssymb}
\usepackage{amsfonts}
\usepackage{amsthm}
\usepackage{amsbsy}
\usepackage{amsgen}
\usepackage{amscd}
\usepackage{amsopn}
\usepackage{amstext}
\usepackage{amsxtra}

\newtheorem{theorem}{Theorem}[section]
\newtheorem{proposition}[theorem]{Proposition}
\newtheorem{lemma}[theorem]{Lemma}
\newtheorem{corollary}[theorem]{Corollary}
\newtheorem{defn}[theorem]{Definition}

\theoremstyle{definition}

\numberwithin{equation}{section}

\newcommand {\Z} {\mathbb{Z}}

\newcommand {\R} {\mathbb{R}}
\newcommand {\T} {\mathbb{T}}

\newcommand {\E} {\mathbb{E}}

\newcommand {\B} {\mathcal {B}}
\newcommand {\C} {\mathbb {C}}

\newcommand{\meas}{\operatorname{meas}}

\newcommand{\Ndim}{\mathcal N}
\newcommand{\eigenvalue}{E}
\newcommand{\eigenspace}{\mathcal{E}}

\newcommand{\TT}{{\mathbb T}}
\newcommand{\var}{\operatorname{Var}}
\newcommand{\tr}{\operatorname{tr}}
\newcommand{\vol}{\operatorname{vol}}
\newcommand{\sing}{B}
\newcommand{\nonsing}{B^c}
\begin{document}

\title[Volume of nodal sets]
{On the volume of nodal sets for eigenfunctions of the Laplacian on
the torus}
\author{Ze\'ev Rudnick and Igor Wigman}
\address{School of Mathematical Sciences, Tel Aviv University,
Tel Aviv 69978, Israel} \email{rudnick@post.tau.ac.il}
\address{Department of Mathematics and Statistics, McGill University and Centre de recherches math\'ematiques (CRM),
Universit\'e de Montr\'eal C.P. 6128, succ. centre-ville Montr\'eal,
Qu\'ebec H3C 3J7, Canada} \email{wigman@crm.umontreal.ca}
\thanks{Z.R. was supported by the Israel Science Foundation
(grant No. 925/06). \\ I.W was supported by CRM analysis laboratory
fellowship}

\date{September 23, 2007}

\begin{abstract}
We study  the volume of nodal sets for  eigenfunctions of the
Laplacian on the standard torus in two or more dimensions.
We consider a sequence of eigenvalues $4\pi^2\eigenvalue$ with growing
multiplicity $\Ndim\to\infty$, and compute the
expectation and variance of the volume of the nodal set with respect
to a Gaussian probability measure on the eigenspaces.
We show that the expected volume of the nodal set is $const
\sqrt{\eigenvalue}$.
Our main result is that the variance of the volume normalized by
$\sqrt{\eigenvalue}$ is bounded by
$O(1/\sqrt{\Ndim})$, so that  the normalized
volume has vanishing fluctuations as we increase the
dimension of the eigenspace.
\end{abstract}
\maketitle

\section{Introduction}
The nodal set of a function on a  manifold is the set of points
where it vanishes. Nodal sets for eigenfunctions of the Laplacian
on a smooth, compact Riemannian manifold have been studied
intensively for some time now. For instance, it is known
\cite{Cheng} that except for a subset of lower dimension, the
nodal sets of eigenfunctions are smooth manifolds of codimension
one in the ambient manifold. In particular one can define their
hypersurface volume (in two dimensions this is the length). A
conjecture of Yau is that the volume of the nodal set is bounded
above and below by constant multiples of square root of the
Laplace eigenvalue. Yau's conjecture was proven for real-analytic
metrics by  Donnelly and Fefferman \cite{Donnelly-Fefferman}. The
lower bound in the case of smooth surfaces is due to Br\"{u}ning
\cite{Bruning}, see also \cite{Bruning-Gromes} for planar domains.

In this paper we study the volume of nodal sets for eigenfunctions
of the Laplacian on the standard flat torus $\TT^d=\R^d/\Z^d$,
$d\geq 2$. We write the eigenvalue equation as $ \Delta
f=-4\pi^2\eigenvalue f$, where $\eigenvalue\geq 0$ is an integer.
The eigenvalues on the torus always have multiplicities, with the
dimension $\Ndim=\Ndim(\eigenvalue)$ of an eigenspace
corresponding to eigenvalue $4\pi^2\eigenvalue$ being the number
of integer vectors $\lambda\in \Z^d$ so that
$|\lambda|^2=\eigenvalue$. In dimension $d\geq5$ this number grows
as $\eigenvalue\to\infty$ roughly as $\eigenvalue^{\frac d2-1}$,
but for small values of $d$, particularly for $d=2$, the behaviour
is more erratic, and depends on the prime decomposition of
$\eigenvalue$.

%As in our paper \cite{ORW},
We will consider {\em random} eigenfunctions
on the torus, that is  random linear combinations
\begin{equation}\label{eq:efn}
f(x)=\frac 1{\sqrt{2\Ndim}} \sum_{\lambda\in \Z^d:
|\lambda|^2=\eigenvalue} b_\lambda \cos 2\pi  \langle \lambda, x
\rangle - c_\lambda \sin 2\pi  \langle \lambda, x \rangle
\end{equation}
with $b_\lambda,c_\lambda\sim N(0, 1)$ real Gaussians of zero mean
and variance $1$ which are independent save for the relations
$b_{-\lambda}=b_\lambda$, $c_{-\lambda}=-c_\lambda$.
Let $\eigenspace=\eigenspace_\eigenvalue$ be the eigenspace
associated to the eigenvalue $4\pi^2 \eigenvalue$ (i.e. the space of
functions of form \eqref{eq:efn}).
We denote by $\E(\bullet)$ the expected value of the quantity $\bullet$
in this ensemble. For instance, the expected amplitude of $f$ is
$\E(|f(x)|^2)=1$.

Denote by $Z(f)$ the volume of the nodal set of an eigenfunction
\eqref{eq:efn}. Our first result, Proposition~\ref{prop:expectation
comp}, is that the expected value of $Z$ is
$$\E(Z) = const\cdot \sqrt{\eigenvalue}$$
for a certain constant depending only on the dimension $d$. This
is of course consistent with the bounds of Donnelly and Fefferman
\cite{Donnelly-Fefferman}.

Our main result, Theorem~\ref{thm:var bnd}, is that
%the normalized volume $Z(f)/\sqrt{\eigenvalue}$
%is a {\em self-averaging} quantity in this ensemble:
%The fluctuations around the mean value die out as the
%multiplicity $\Ndim\to\infty$. Precisely, we show that
the variance of the normalized volume $Z/\sqrt{\eigenvalue}$ is bounded by
$$\var(\frac{Z}{\sqrt{\eigenvalue}})\ll \frac 1{\sqrt{\Ndim}}, \qquad
\mbox{as }\Ndim\to \infty\;.
$$
(We believe that the correct upper bound for the variance is
$O(1/\Ndim)$). Thus the fluctuations of $Z(f)/\sqrt{\eigenvalue}$
around its mean value die out as the multiplicity $\Ndim$ tends to
$\infty$. Note however that $Z(f)/\sqrt{\eigenvalue}$ is not
asymptotically constant; for instance, if $\eigenvalue =dm^2$ then
for the eigenfunction $f(x)=\prod_{j=1}^d \sin 2\pi m x_j$ we have
$Z(f)/\sqrt{\eigenvalue}=2\sqrt{d}$ while if $\eigenvalue = m^2$
then for the eigenfunction $f(x)=\sin2\pi mx_1$ we have
$Z(f)/\sqrt{\eigenvalue} = 2$.

Theorem~\ref{thm:var bnd} can be viewed as lending support to the
expectation\footnote{We thank Steve Zelditch for a discussion of
this.} that for eigenfunctions on negatively curved manifolds,
which are believed to behave similarly to random waves \cite{Berry
1977}, the volumes of nodal sets, normalized by the square-root of
the eigenvalue, do tend to a limiting value. See
\cite{Zelditch-complex} for some work on the {\em complexified}
nodal set of eigenfunctions in this context.

Previous work in this vein is due to B\'erard \cite{Berard}, who
computed the expected surface measure of the nodal set for
eigenfunctions of the Laplacian on spheres. Neuheisel
\cite{Neuheisel} also worked on the sphere and  gave an upper bound
for the variance. Berry \cite{Berry 2002} computed the expected
length of nodal lines for isotropic, monochromatic random waves in
the plane, which are eigenfunctions of the Laplacian with eigenvalue
$\eigenvalue$. He found that the expected length (per unit area) is
again of size about $\sqrt{\eigenvalue}$ and argued that the
variance should be of order $\log \eigenvalue$.

%Other characteristics of the nodal set of eigenfunctions have been
%studied intensively in the last few years. We mention the work on
%nodal domains \cite{Blum, Bogomolny},   area ,
%extrema on nodal domains \cite{Polterovich-Sodin extrema}
%For some work on the {\em complexified}  nodal set of eigenfunctions,
%see \cite{Zelditch-complex}.

More recently, F.~Oravecz and ourselves have investigated a different
characteristic of the nodal set of eigenfunctions on the torus, namely
the Leray nodal measure \cite{ORW},
and have succeeded in obtaining the precise
asymptotics of the variance of the Leray measure as $\Ndim\to
\infty$.
%In this work we will need to use several results obtained in
%the paper \cite{ORW}.

\subsection{Plan of the paper}
We employ a version of the Kac-Rice formula for the volume of the nodal set,
which using the Dirac delta function can be written as
$$ Z(f) = \int_{\TT^d} \delta(f(x)) |\nabla f(x)| dx \;,$$
see \S\ref{sec:volume} for the rigorous version. To compute the
expected value of $Z$ is then a simple matter once we find that
$f(x)$ and $\partial f/\partial x_j$ are independent Gaussians.
This is done in \S~\ref{sec:expectation}. In \S~\ref{sec:second
moment} we  derive a formula for the second moment of $Z$, which
requires knowing the covariance structure of the
$2d+2$-dimensional Gaussian vector $v(x,y)=(f(x),f(y),\nabla
f(x),\nabla f(y))$. That $v(x,y)$ is indeed a non-degenerate
$2d+2$ dimensional Gaussian is verified in the Appendix.   As a
result, we find that $\E(Z^2)  = \int_{\TT^d} K(z)dz$, with
$$
K(z)=\frac 1 {\sqrt{1-u(z)^2}} \int_{\R^{2d}}\ \| v_1\| \| v_2\|
\frac {\exp(-\frac 12  v\Omega(z)^{-1}  v^T)} {\sqrt{\det\Omega(z)}}
\frac{d v}{(2\pi)^{d+1}}\;,
$$
where $u(z) = \E(f(x) f(x+z))$ is the two-point function of the
ensemble, and where $\Omega(z)$ is a certain positive definite
$2d\times 2d$ matrix which enters into the covariance structure of
the Gaussian vector $v(x,y)$. In section \S\ref{sec:variance},
which is the heart of the paper, we bound the variance of $Z$.

\section{The model: random eigenfunctions on the torus}
\subsection{Random eigenfunctions}
We consider non-constant eigenfunctions of the Laplacian on the standard
flat torus $\TT^d=\R^d/\Z^d$. The solutions of the eigenvalue equation
$$\Delta\psi+4\pi^2\eigenvalue\psi=0\;, \qquad \eigenvalue\neq 0\;,
$$
form a finite dimensional vector space $\eigenspace =
\eigenspace_\eigenvalue$, having as a basis the
exponentials $e^{2\pi i \langle \lambda,x \rangle}$, for $\lambda$ in
the frequency set
$$\Lambda=\Lambda_\eigenvalue=\{\lambda\in \Z^d,
|\lambda|^2=\eigenvalue\}\;.
$$

We define an ensemble of Gaussian random
functions $f\in\eigenspace$ by
\begin{equation*}
%\label{eq:gen def f}
f(x)=\frac 1{\sqrt{2\Ndim}} \sum_{\lambda\in \Lambda} b_\lambda
\cos 2\pi  \langle \lambda, x \rangle - c_\lambda \sin 2\pi
\langle \lambda, x \rangle
\end{equation*}
with $b_\lambda,c_\lambda\sim N(0, 1)$ real  Gaussians of zero
mean and variance $1$ which are independent save for the relations
$b_{-\lambda}=b_\lambda$, $c_{-\lambda}=-c_\lambda$.
Thus we can rewrite
\begin{equation}\label{eq:second expr for f}
f(x)=\sqrt{\frac 2 {\Ndim}} \sum_{\lambda\in \Lambda/\pm}
b_\lambda \cos 2\pi  \langle \lambda, x \rangle - c_\lambda \sin
2\pi  \langle \lambda, x \rangle
\end{equation}
where now only independent random variables appear.  With our
normalization,  we have $\E(|f(x)|^2)=1$ for all $x\in \TT^d$.

%\subsection{Singular eigenfunctions}
\begin{defn}
An eigenfunction $f\in \eigenspace$ is {\em singular}  if
$\exists x\in \T^d$ with  $f(x)=0$  and $\nabla f(x) = \vec{0}$.
An eigenfunction $f\in \eigenspace$ is  {\em nonsingular}
if $\nabla f\neq \vec 0$ on the nodal set.
\end{defn}

\begin{lemma}[\cite{ORW}, Lemma 2.3]
\label{lem:sing codim 1}
The set of singular eigenfunctions  has codimension at least
$1$ in $\eigenspace$, and so has measure zero in $\eigenspace$.
\end{lemma}

\subsection{Properties of the frequency set}
The dimension $\Ndim = \dim \eigenspace$ is the number of the frequencies in
$\Lambda$, which
%The dimension $\Ndim=\#\Lambda$ %of the  eigenspace
is  the number  of ways of expressing $\eigenvalue$ as a sum of $d$
integer squares.
For $d\geq 5$ this grows roughly as  $\eigenvalue^{d/2-1}$ as
$\eigenvalue\to\infty$. For $d\leq 4$ the  dimension of the eigenspace
need not grow with $\eigenvalue$. For instance, for $d=2$,
$\Ndim$ is given in terms of the prime
decomposition of $\eigenvalue$ as follows:
If $\eigenvalue=2^\alpha\prod_j p_j^{\beta_j}\prod_k
q_k^{2\gamma_k}$  where $p_j\equiv 1 \mod 4$ and $q_k\equiv 3\mod 4$
are odd primes, $\alpha,\beta_j,\gamma_k\geq$ are integers,
then  $\Ndim = 4\prod_j(\beta_j+1)$, and otherwise $\eigenvalue$ is not
a sum of two squares and $\Ndim=0$. On average (over integers which are
sums of two squares) the dimension is $const\cdot \sqrt{\log \eigenvalue}$.

The frequency set $\Lambda$ is invariant under the group $W_d$
of signed permutations, consisting of
coordinate permutations and sign-change of any coordinate,
e.g. $(\lambda_1,\lambda_2)\mapsto (-\lambda_1,\lambda_2)$ (for
$d=2$). In particular $\Lambda$ is symmetric under $\lambda\mapsto
-\lambda$ and since $0\notin \Lambda$, we find $\Ndim$ is even.
We write $\Lambda/\pm$ to denote representatives of
the equivalence class of $\Lambda$ under $\lambda\mapsto -\lambda$.

We will need some simple properties of $\Lambda$:

\begin{lemma}
 For any subset $\mathcal O\subset \Lambda$ which is invariant
under the group $W_d$, we have
\begin{equation}\label{eq:identity for lambdas}
\frac{1}{|\mathcal O|}\sum_{\lambda\in\mathcal O} \lambda_j
\lambda_k  = \frac{\eigenvalue}{d} \cdot\delta_{j,k} \;.
\end{equation}
Moreover for any $C\in \R^d$,
\begin{equation}\label{ORW, Lemma 5.2}
\frac 1{|\mathcal O|}\sum_{\lambda\in \mathcal O} \langle
 C,\lambda\rangle^2  =\frac {\eigenvalue}d  |C|^2 \;.
\end{equation}
\end{lemma}
\begin{proof}
 For $i\neq j$ use the symmetry of $\mathcal O$
%the frequency set $\Lambda$
under the sign change of the
$i$-th coordinate to change variables and deduce that the LHS of
\eqref{eq:identity for lambdas} vanishes.
For $i=j$ note that the sum $ \sum_{\lambda\in \mathcal O} \lambda_i^2$
is independent of $i$ since $\mathcal O$ is
symmetric under permutations; hence we may average the RHS over
$i$ to find that
$$
\sum_{\lambda\in \mathcal O} \lambda_i^2=
\frac 1d \sum_{i=1}^d \sum_{\lambda\in \mathcal O} \lambda_i^2 =
\frac 1d \sum_{\lambda\in \mathcal O} ||\lambda||^2 =
\frac{|\mathcal O|\eigenvalue}{d}
\;,
$$
proving \eqref{eq:identity for lambdas}.
To prove \eqref{ORW, Lemma 5.2} we expand
$\langle C,\lambda\rangle^2 = \sum_{j,k=1}^d c_jc_k\lambda_j\lambda_k$
and use \eqref{eq:identity for lambdas}.
\end{proof}
Note that \eqref{ORW, Lemma 5.2} implies  that the frequency set
$\Lambda$ spans $\R^d$.  %(This is proved in  \cite[Lemma 2.1]{ORW}).

\subsection{The two point function}
The two-point function of the ensemble is
\begin{equation}\label{eq:two point function}
u(z):=\E(f(x+z)f(x)) = \frac{2}{\Ndim} \sum_{\lambda\in
\Lambda/\pm} \cos  2\pi \langle  \lambda ,z \rangle   \;.
\end{equation}
The two-point function clearly satisfies $|u(z)|\leq 1$.
We will need to know some of its basic properties, proved in
\cite{ORW}, which we summarize as:
\begin{proposition}\label{prop:two point function}
The two point function satisfies
\begin{enumerate}
\item \label{item 1} 
There are only finitely many points $x\in \TT^d$ where $u(x)=\pm 1$.

\item \label{item 2} 
The mean square of $u$ is 
$%\begin{equation*}
\int_{\TT^d} u^2= \frac 1{\Ndim} \;.
$%\end{equation*}
%(as follows from the formula~\eqref{eq:two point function}).

\item\label{item 3} 
The mean fourth power of $u$ is bounded by\footnote{Except possibly
in dimensions $d=3,4$ we have a better bound in \cite{ORW} of
$o(1/\Ndim)$, though we have no use for this finer information in
this paper.}
%\begin{equation*}
$\int_{\TT^d} u^4 \ll 1/\Ndim \;.$
%\end{equation*}

\item\label{item 4}
The kernel $1/\sqrt{1-u^2}$ is integrable on $\TT^d$.
\end{enumerate}
\end{proposition}
Part~\ref{item 1} follows from  \cite[lemma 2.2]{ORW}, 
part~\ref{item 3} is \cite[Proposition 7.1]{ORW}, 
and part~\ref{item 4} is \cite[Lemma 5.3]{ORW}.

\section{A formula for the volume of the nodal set}
\label{sec:volume}

Let $\chi$ be the indicator function of the interval $[-1,1]$. We
define for $\epsilon>0$
$$
Z_\epsilon(f):=\frac{1}{2\epsilon} \int_{\T^d}
\chi\bigg(\frac{f(x)}\epsilon\bigg) |\nabla f(x)|dx \;.
$$

\begin{lemma}\label{lem: formula for Z}
Suppose that $f\in \eigenspace$ is non-singular. Then
$$ \vol(f^{-1}(0))=\lim_{\epsilon \to 0} Z_\epsilon(f) \;.$$
\end{lemma}
\begin{proof}
By the co-area formula \cite{Federer}, for $f$ smooth and
$\phi$ integrable, we have
$$\int_{\TT^d} \phi(x) |\nabla f(x)|dx = \int_{-\infty}^{\infty} \left(
\int_{f^{-1}(s) } \phi(x) dx \right) ds \;.
$$
Taking $\phi(x):=\frac 1{2\epsilon} \chi(\frac{f(x)}{\epsilon})$, which
is constant on the level sets $f^{-1}(s)$ gives
$$Z_\epsilon(f) = \frac 1{2\epsilon}\int_{-\epsilon}^{\epsilon}
\vol(f^{-1}(s)) ds \;.
$$
Now if $f$ is non-singular then $s\mapsto \vol(f^{-1}(s)) $ is
continuous at $s=0$ and so by the fundamental theorem of calculus,
$$ \lim_{\epsilon\to 0} \frac 1{2\epsilon}\int_{-\epsilon}^{\epsilon}
\vol(f^{-1}(s)) ds = \vol(f^{-1}(0)) \;.
$$
Thus $\lim_{\epsilon\to 0} Z_\epsilon(f) =\vol(f^{-1}(0))$ as
claimed.
\end{proof}

\begin{lemma}\label{lem:uniform bd on Z_epsilon}
For all $f\in \eigenspace$
%(that is, functions of form \eqref{eq:efn})
we have
$$Z_\epsilon(f)\leq 6d \sqrt{\eigenvalue}\;.$$
\end{lemma}

We begin with the one variable case which we state as a separate
lemma (cf \cite[Lemma 2]{Kac}):
\begin{lemma}\label{lem:Kaclength}
Let $g(t)$
be a trigonometric polynomial of degree at most $M$. Then for all
$\epsilon>0$ we have
$$
\frac 1 {2\epsilon}\int_{\{t: |g(t)|\leq \epsilon \}}|g'(t)|dt \leq
6M \;.
$$
\end{lemma}
\begin{proof}
We partition the set $\{t:|g(t)|\leq  \epsilon \}\subseteq [0,1]$
into a union of maximal closed intervals $[a_k,b_k]$ (with
$a_k<b_k$), disjoint except perhaps for common edges,  such that on
each such interval $g'$ has constant sign, that is either $g'\geq 0$
or $g'\leq 0$. If $g'\geq 0$ on $[a_k,b_k]$ then either
$g(a_k)=-\epsilon$  or $g'(a_k)=0$ and $a_k$ is a local minimum for
$g$, and $g(b_k)\leq +\epsilon$. If $g'\leq 0$ on $[a_k,b_k]$ then
either $g(a_k)=+\epsilon$  or $g'(a_k)=0$ and $a_k$ is a local
maximum for $g$, and $g(b_k) \geq -\epsilon$.

If $g'\geq 0$ on $[a_k,b_k]$ then
$$\int_{a_k}^{b_k} |g'(t)|dt = \int_{a_k}^{b_k} g'(t)dt =
g(b_k)-g(a_k) \leq 2\epsilon\;,
$$
while if $g'\leq 0$ on $[a_k,b_k]$ then
$$
\int_{a_k}^{b_k} |g'(t)|dt = \int_{a_k}^{b_k} -g'(t)dt =
g(a_k)-g(b_k) \leq 2\epsilon \;.
$$

Thus the total integral is bounded by the number $\nu$ of intervals
$[a_k,b_k]$:
$$
\frac 1 {2\epsilon}\int_{\{t: |g(t)|\leq \epsilon\} }|g'(t)|dt \leq
\nu \;.
$$
Now the number of intervals is bounded by the number of $a$'s for
which $g(a)=\pm \epsilon$ plus the number of $a$'s for which
$g'(a)=0$. Since both $g$ and $g'$ are trigonometric polynomials of
degree $\leq M$, the number of such intervals is therefore $3\cdot
2M=6M$. This gives the required bound.
\end{proof}
We now prove Lemma~\ref{lem:uniform bd on Z_epsilon} by reduction to
the one-dimensional case.
\begin{proof}
Since $|\nabla f| \leq \sum_{j=1}^d |\frac{\partial f}{\partial
x_j}|$ we have
$$
Z_\epsilon(f) \leq \sum_{j=1}^d \frac 1{2\epsilon} \int_{\TT^d}
\chi(\frac{f(x)}\epsilon)|\frac{\partial f}{\partial x_j}| dx
$$
and we will bound each term. Taking $j=1$, we have
$$
\frac 1{2\epsilon} \int_{\TT^d}
\chi\bigg(\frac{f(x)}\epsilon\bigg) \bigg| \frac{\partial
f}{\partial x_1}\bigg| dx = \int_{\vec y\in \TT^{d-1}} \left(
\frac 1{2\epsilon} \int_{\{t\in \TT^1 : |f(t,\vec y)|\leq \epsilon
\}} \bigg|\frac{\partial f(t,\vec y)}{\partial t}\bigg| dt \right)
d\vec y \;.
$$
In the inner integral we have for each $\vec y\in \TT^{d-1}$ a one
variable polynomial  $g(t)=f(t,\vec y)$ of degree at most
$\sqrt{\eigenvalue}$ and hence by Lemma~\ref{lem:Kaclength}, the
inner integral is at most $ 6\sqrt{\eigenvalue}$. Summing over $j$
introduces another factor of $d$.
\end{proof}

%\subsection{Computing moments of $Z$}
As a consequence of the fact that for nonsingular functions we can
compute the volume $Z(f)$ of the nodal set of $f\in \eigenspace$
via Lemma~\ref{lem: formula for Z} and the fact
that almost all $f\in \eigenspace$ are nonsingular
(Lemma~\ref{lem:sing codim 1}), we find:
\begin{corollary} \label{cor:formulas for moments}
The first and second moments of the volume $Z(f)$ of the nodal set of
$f$ are given by
$$\E(Z) = \E(\lim_{\epsilon\to 0} Z_\epsilon),\qquad
\E(Z^2) = \E(\lim_{\epsilon_1,\epsilon_2\to 0} Z_{\epsilon_1}
Z_{\epsilon_2}) \;.
$$
\end{corollary}

\section{The expected volume of the nodal set}
\label{sec:expectation} In this section we show
\begin{proposition}
\label{prop:expectation comp} For $d\geq 1$,
\begin{equation*}
\label{eq:expectation comp} \E(Z) = \mathcal I_d \sqrt{\eigenvalue}
\end{equation*}
where $$\mathcal I_d =
\sqrt{\frac{4\pi}{d}}\frac{\Gamma(\frac{d+1}2)}{\Gamma(\frac d2)}
\;.$$
\end{proposition}

\begin{proof}
Since $Z_\epsilon$ is uniformly bounded by Lemma~\ref{lem:uniform bd
on Z_epsilon}, we can use the Dominated Convergence Theorem to write
$$
\E(Z) = \E(\lim_{\epsilon\to 0} Z_\epsilon) = \lim_{\epsilon\to 0}
\E(Z_\epsilon) \;.
$$
By Fubini's theorem,
\begin{multline*}
\E(Z_\epsilon) = \E\left( \frac 1{2\epsilon} \int_{\TT^d}
\chi(\frac{f(x)}{\epsilon}) |\nabla f(x)|dx \right) \\
= \int_{\TT^d}  \E\left( \frac 1{2\epsilon}
\chi(\frac{f(x)}{\epsilon}) |\nabla f(x)|\right) dx =:
\int_{\TT^d} K_\epsilon(x)dx \;.
\end{multline*}
Now for each $x\in \TT^d$, the function $f\mapsto \frac 1{2\epsilon}
\chi(\frac{f(x)}{\epsilon}) |\nabla f(x)|$ is bounded
and hence we may evaluate the integral by using the joint
probability density of the variable $(f(x),\nabla f(x))$, whose
components are Gaussian of zero mean with covariance
$$
\E(f(x)^2) = 1,\qquad \E\bigg(f(x)\frac{\partial f}{\partial
x_j}(x)\bigg) = 0
$$
and

\begin{equation}\label{eq:covariances of f}
\E(\frac{\partial f(x)}{\partial x_j}\frac{\partial f(x)}{\partial
x_k})= \frac{2}{\Ndim}\cdot 4\pi^2 \sum_{\vec \lambda\in
\Lambda/\pm} \lambda_j \lambda_k = 4\pi^2\frac{\eigenvalue}{d} \cdot
\delta_{j,k}.
\end{equation}
by \eqref{eq:identity for lambdas}.
Thus
\begin{equation*}
\begin{split}
K_\epsilon(x) &= \frac 1{2\epsilon}\int_{\R} \chi(\frac a\epsilon)
e^{-a^2/2} \frac{da}{\sqrt{2\pi}}
\int_{\R^d}  |\vec b|\exp(-\frac{d|\vec b|^2 }{8\pi^2\eigenvalue})
\frac{d\vec b}{(2\pi)^{d/2}(4\pi^2\eigenvalue/d)^{d/2}} \\
&= \frac{\sqrt{4\pi^2\eigenvalue}}{\sqrt{d}\cdot(2\pi)^{(d+1)/2}}
\int_{\R^{d}} |\vec v| \exp(-\frac 12 |\vec v|^2) d\vec v \frac
1{2\epsilon}\int_{\R} \chi(\frac a\epsilon) e^{-a^2/2} da \;.
\end{split}
\end{equation*}
Integrating over $\TT^d$ and taking the limit $\epsilon\to 0$ gives
$$\E(Z) = \mathcal I_d \sqrt{\eigenvalue}
$$
where
$$\mathcal I_d = \frac{1}{\sqrt{d}(2\pi)^{(d-1)/2 }}
\int_{\R^{d}} |\vec v| \exp(-\frac 12 |\vec v|^2) d\vec v \;.
$$

In the one-dimensional  case, $\mathcal I_1=\int_\R
|v|e^{-v^2/2}dv = 2$. For $d\geq 2$,
\begin{equation*}
\int_{\R^{d}}  |\vec v|\exp(-\frac 12 |\vec v|^2) d\vec v =
vol(S^{d-1})\int_0^\infty r e^{-r^2/2} r^{d-1}dr \;.
\end{equation*}
Using
$$
vol(S^{d-1}) = \frac{2 \pi^{\frac{d}2}}{\Gamma(\frac{d}2)}, \qquad
\int_0^\infty r^de^{-r^2/2} dr = 2^{\frac{d-1} 2}
\Gamma(\frac{d+1}2)
$$
gives
\begin{equation*}
\int_{\R^{d}}  |\vec v|\exp(-\frac 12 |\vec v|^2) d\vec v
= \sqrt{2} (2\pi)^{d/2}\frac{\Gamma(\frac{d+1}2)}{\Gamma(\frac d2)}
\end{equation*}
(which is consistent with the computation for $d=1$). Thus
$$
\mathcal I_d=
\sqrt{\frac{4\pi}{d}}\frac{\Gamma(\frac{d+1}2)}{\Gamma(\frac d2)}
$$
as claimed.
\end{proof}

\section{An integral formula for the second moment}
\label{sec:second moment}

\subsection{The covariance matrix}
The covariance matrix $\Sigma(x,y)$ of the Gaussian vector
$(f(x),f(y),\nabla f(x), \nabla f(y))$  is given by
\begin{equation}
\label{eq:Sigma def} \Sigma =
\begin{pmatrix}
      A&B\\B^T&C
      \end{pmatrix}
\end{equation}
where
$$
A%=\E(f(x,y)^T f(x,y))
= \begin{pmatrix}  \E(f(x)^2)&\E(f(x)f(y))\\ \E(f(y)f(x))&\E(f(y)^2)
\end{pmatrix}
,\;
B%=\E(f(x,y)^T D(x,y))
=\begin{pmatrix} \E(f(x)\nabla f(x))&\E(f(x)\nabla f(y))\\
\E(f(y)\nabla f(x)&\E(f(y)\nabla f(y))
           \end{pmatrix}
$$
and
$$ C%=\E(D(x,y)^T D(x,y))
= \begin{pmatrix}   \E(\nabla f(x)^T \nabla f(x))&\E(\nabla f(x)^T \nabla f(y))\\
\E(\nabla f(y)^T \nabla f(x)) & \E(\nabla f(y)^T \nabla f(y))
\end{pmatrix} \;.
$$
For generic $(x,y)$, the covariance matrix $\Sigma(x,y)$ is
nonsingular (see Appendix~\ref{sec:nondeg of cov}).

\begin{lemma}\label{prop covariance}
The covariance matrix $\Sigma(x,y)$ depends only on the difference
$z=x-y$ and is given in terms of the two-point function $u$ by
$$
\Sigma(x,y) = \begin{pmatrix}
     A(z)&B(z)\\B(z)^T&C(z)
     \end{pmatrix}
$$
where
$$
A(z)=\begin{pmatrix}    1&u(z)\\u(z)&1    \end{pmatrix},\qquad
B(z)=\begin{pmatrix}
   \vec 0&-\nabla u(z) \\ \nabla u(z)&\vec 0
  \end{pmatrix}
$$
(here $\vec 0$,$\nabla u$ are row vectors), and
$$
C(z)=\begin{pmatrix}
\frac{4\pi^2 \eigenvalue}{d} I& -H(z)\\
-H(z)&  \frac{4\pi^2 \eigenvalue}{d}I
\end{pmatrix}
$$
where $H=(\frac{\partial^2 u}{\partial x_j\partial x_k})$ is the
Hessian of $u$.
\end{lemma}

\begin{proof}
By definition of the two point function, we have $A=\begin{pmatrix}
1&u\\u&1 \end{pmatrix}$.
To compute $B$, use
$$\E(\frac{\partial f}{\partial x_j}(x) f(y)) =
\frac{\partial}{\partial x_j} \E(f(x) f(y)) = \partial_j u(x-y)
$$
and hence
$$
\E(f(x)\partial_j f(y)) = \partial_j u(y-x) = -\partial_j
u(x-y)\;.
$$
In particular
$$
\E(f(x)\nabla f(x)) = \vec 0
$$
Therefore
$$
B(z)=\begin{pmatrix}
   \vec 0&-\nabla u(z) \\ \nabla u(z)&\vec 0
  \end{pmatrix}
$$
(where $\vec 0$ denotes the $d$-dimensional zero row vector).

To compute $C$, use \eqref{eq:covariances of f} to find
$$
\E(\nabla f(x)^T \nabla f(x)) = \frac{4\pi^2 \eigenvalue}{d}  I_d
\;.
$$
More generally
$$\E(\partial_j f(x)\partial_k f(y)) =
\frac{\partial^2 }{\partial x_j\partial y_k} \E(f(x)f(y)) =
-\frac{\partial^2 u}{\partial x_j\partial x_k}(x-y)
$$
and so
$$\E(\nabla f(x)^T \nabla f(y)) =
-(\frac{\partial^2 u}{\partial x_j\partial x_k}(x-y))_{j,k} =-H(x-y) \;.
$$
Thus
$$
C=\begin{pmatrix}  \frac{4\pi^2 \eigenvalue}{d}  I& -H\\
-H& \frac{4\pi^2 \eigenvalue}{d}  I
\end{pmatrix}
$$
as claimed.
\end{proof}
The inverse of $\Sigma$ (when it exists) is given by
$$
\Sigma^{-1} = \begin{pmatrix} *&*\\ *& \Omega^{-1} \end{pmatrix}
$$
with $\Omega$ being the $2d\times 2d$ matrix
$$
\Omega = C-B^TA^{-1}B\;.
$$
We will call $\Omega$ the {\em reduced covariance matrix}.
We have %By Jacobi's formula,
\begin{equation}
\label{eq:Jacobi's formula}
\det{\Sigma}= \det{A}\det{\Omega} = (1-u^2)\det{\Omega} \;.
\end{equation}

By Lemma~\ref{prop covariance}, we have
\begin{equation}\label{expression for Omega}
\begin{split}
\Omega=\begin{pmatrix}  4\pi^2 (\eigenvalue/d)I &-H\\ -H& 4\pi^2
(\eigenvalue/d) I
\end{pmatrix} - \frac 1{1-u^2} \begin{pmatrix}  D^T D& u D^TD\\ uD^T
D& D^TD
\end{pmatrix}
\end{split}
\end{equation}
where $D(z) = \nabla u(z)$ and $H=(\frac{\partial^2 u}{\partial x_j
x_k})$ is the Hessian of $u$.

\subsection{A formula for the second moment}
\begin{proposition}
\label{prop:justif ord chng}
The second moment of $Z(f)$ is given by
\begin{equation}
\label{eq:int form sec mom}
\E(Z^2) = \int_{\T^d} K(x) dx
\end{equation} where
\begin{equation}\label{eq:formula for K(x)}
K(x) = \frac 1 {\sqrt{1-u^2}} \int_{\R^{2d}}\ \| v_1\| \| v_2\|
\frac {\exp(-\frac 12  v\Omega^{-1}  v^T)} {\sqrt{\det\Omega}}
\frac{d v}{(2\pi)^{d+1}}\;.
\end{equation}
\end{proposition}

Denote
\begin{equation*}
\label{eq:def K eps(x,y)} K_{\epsilon_1, \epsilon_2} (x,y) :=
\frac{1}{4\epsilon_1\epsilon_2} \int_{\eigenspace} \|\nabla f(x)
\| \|\nabla f(y) \| \chi \bigg( \frac{f(x)}{\epsilon_1} \bigg)
\chi \bigg( \frac{f(y)}{\epsilon_2} \bigg) d\mu(f) \;.
\end{equation*}

We have the following
\begin{lemma}\label{lem:bound on K eps(x,y)}
For $(x,\, y)\in\T^{d}\times\T^{d}$ with $u(x-y)^2 \ne 1$
\begin{equation}
\label{eq:bnd krnl eps}
K_{\epsilon_1,\, \epsilon_2} (x,\, y)\ll_d
\frac{\eigenvalue}{\sqrt{1-u^2(x-y)}},
\end{equation}
where the implied constant depends only on the dimension $d$.
\end{lemma}

\begin{proof}%[Proof of lemma \eqref{lem:bound on K eps(x,y)}]
Write $f(x) = \langle f,\, U(x) \rangle$, where $U(x)$ is the unit
vector $$U(x) = \frac{\sqrt{2}}{\sqrt{\Ndim}}
\bigg(\cos 2\pi \langle \lambda, x \rangle ,
\sin 2\pi\langle \lambda, x \rangle \bigg)_{\lambda \in\Lambda/\pm}
\in S^{\Ndim-1},
$$
and where we identify the function $f$ with a vector in $\R^\Ndim$
via \eqref{eq:second expr for f}.
Note that $\langle U(x),U(y) \rangle = u(x-y)$ is the cosine of the
angle between $U(x)$ and $U(y)$.

We have
$$\nabla f(x) =  DU \cdot f$$
where the derivative $DU$
is a $d\times \Ndim$ matrix. Equivalently,
$$\big(\nabla f(x)\big)_i =
\bigg\langle f, \bigg(\frac{\partial}{\partial x_i} U(x) \bigg)
\bigg\rangle ,\quad 1\le i\le d.$$

By the triangle and Cauchy-Schwartz inequalities, $$\|\nabla f (x)
\| \le \sum_{i=1}^{d} \| f\|\cdot \bigg\|
\bigg(\frac{\partial}{\partial x_i} U(x) \bigg) \bigg\| \ll
\sqrt{\eigenvalue} \| f \|,
$$
by a computation of $\frac{\partial U}{\partial x_i}$.
Therefore
\begin{equation}
\label{eq:mult int Keps}
K_{\epsilon_1,\epsilon_2} (x,y) \ll
\frac{\eigenvalue}{4\epsilon_1
\epsilon_2}\int_{\substack{|f(x)| < \epsilon_1 \\
|f(y)|<\epsilon_2}} \|f\|^2 e^{-\| f \| ^2/2} df \;.
\end{equation}

Consider the plane $\pi\subset \R^{\Ndim}$ spanned by $U(x)$ and
$U(y)$. The domain of the integration is all the vectors
$f\in\R^\Ndim$ so that the projection of $f$ on $\pi$ falls into the
parallelogram $P$ of lengths $2\epsilon_1$ and $2\epsilon_2$. The
cosine of the angle  between the sides of $P$ is
$\langle U(x), \,U(y) \rangle = u(x-y)$.
Therefore the area of $P$ is
$$
\mbox{area}(P) = 4\epsilon_1 \epsilon_2 \frac{1}{\sqrt{1-u(x-y)^2}} \;.
$$
Write the multiple integral in \eqref{eq:mult int Keps} as the
iterated integral
\begin{equation}
\label{eq:mult int K rep} \int_{P}
\left(\int_{p+\pi^{\perp}} \| f\|^2 e^{-\| f \|^2/2} df\right) dp \;,
\end{equation}
where the variable $p$ runs over all the points of the parallelepiped
$P$.
The inner integral in \eqref{eq:mult int K rep} is $O(1)$ with the
constant depending on $d$ only. Indeed, note that for every
$f_1\in\pi^{\perp}$,
\begin{equation*}
\begin{split}
\|
p+f_1\|^2 e^{-\|p+f_1\| ^2/2}
&= (\|p\|^2+\|f_1\|^2)e^{-(\|p\|^2+\|f_1\| ^2)/2} \\
&\ll (1+\|f_1\|^2) \cdot e^{-\|f_1\| ^2/2}\;,
\end{split}
\end{equation*}
since $\|p\|^2 e^{-\| p \| ^2/2}$ is bounded.
Our claim follows from convergence of the integral
$\int_{\R^{\Ndim-2}} (1+\|w\|^2)e^{-\| w\|^2/2}dw $.
Therefore
$$
\int_{\substack{|f(x)| < \epsilon_1 \\
|f(y)|<\epsilon_2}} \|f\|^2 e^{-\| f \| ^2/2} df \ll area(P) \ll
\epsilon_1\epsilon_2 \frac{1}{\sqrt{1-u(x-y)^2}} \;.
$$
Substituting the last estimate into \eqref{eq:mult int Keps} proves
\eqref{eq:bnd krnl eps}.
\end{proof}

\begin{proof}[Proof of proposition \ref{prop:justif ord chng}]
By Corollary~\ref{cor:formulas for moments}, we have
\begin{multline*}
\E(Z^2) =\\
\int_{\eigenspace}
\bigg[ \lim_{\epsilon_1,\epsilon_2 \to 0} \frac{1}{2\epsilon_1}
\int_{\T^d} \|\nabla f(x) \| \chi(\frac{f(x)}{\epsilon_1} ) dx
\frac{1}{2\epsilon_2} \int_{\T^d} \|\nabla f(y) \|
\chi( \frac{f(y)}{\epsilon_2}) dy\bigg] d\mu(f)
\end{multline*}
where
$\mu$ is the Gaussian measure $d\mu(f) = e^{-\| f\|^2/2}
\frac{df}{(2\pi)^{\Ndim/2}}$.
We wish to change the order of the limit and the integration.
To do so, we notice that by Lemma~\ref{lem:uniform bd on Z_epsilon},
the integrand is bounded by $O(\eigenvalue)$. Therefore, the change of
order follows from the dominated convergence theorem. Thus the
integral equals
$$\lim_{\epsilon_1, \epsilon_2\to  0}
\frac{1}{4\epsilon_1\epsilon_2} \int_{\eigenspace}
\int_{(\T^d)^2} \|\nabla f(x) \| \|\nabla f(y) \| \chi \bigg(
\frac{f(x)}{\epsilon_1} \bigg) \chi \bigg( \frac{f(y)}{\epsilon_2}
\bigg) dx dy  d\mu(f)\;.
$$
Using Fubini's theorem, this equals to
\begin{equation}
\label{eq:exchng ord stp2}
\lim_{\epsilon_1,\epsilon_2\to 0}
\frac{1}{4\epsilon_1\epsilon_2} \int_{(\T^d)^2} \int_{\eigenspace}
\|\nabla f(x) \| \|\nabla f(y) \|
\chi \bigg( \frac{f(x)}{\epsilon_1} \bigg) \chi \bigg(
\frac{f(y)}{\epsilon_2} \bigg) d\mu(f) dx dy \;.
\end{equation}

Now we wish to exchange the order of taking limit and the
integration over $(\T^d)^2$. To justify it, we use the dominated
convergence theorem with Lemma~\ref{lem:bound on K eps(x,y)}. The
upper bound for $u(x)\ne \pm u(y)$ is sufficient, since this happens
for almost all $(x,y)\in (\T^d)^2$, and changing the values of a
function on a set of measure $0$ does not have any impact on the
integrability and the value of the integral of a function. The
convergence of the RHS of \eqref{eq:bnd krnl eps} was shown in
\cite{ORW}. Therefore, we may exchange the
order of the limit and the integral in \eqref{eq:exchng ord stp2} to
obtain
\begin{equation}
\label{eq:int form sec mom dbl}
\E(Z^2) = \int_{\TT^d\times \TT^d} K(x, y)dxdy,
\end{equation}
where
$$
K(x,y) = \lim_{\epsilon_1,\epsilon_2\to 0 }
K_{\epsilon_1,\epsilon_2}(x,y)\;.
$$

We will replace the  vector $(f(x),f(y),\nabla f(x), \nabla f(y))$
with a $2d+2$ dimensional Gaussian vector with covariance matrix
$\Sigma(x,y)= \Sigma(x-y)$ defined in \eqref{eq:Sigma def}. The
proof that for almost all $x,y$ this  is indeed a $2d+2$
dimensional process, is relegated to Proposition~\ref{prop:gen
nondeg proc} in the Appendix. This gives
\begin{equation}\label{eq: change vars}
\begin{split}
&K_{\epsilon_1,\epsilon_2}(x,y) \\
&=
\frac{1}{4\epsilon_1\epsilon_2} \int_{\R^{2d+2}} \|v_1\|
\|v_2\| \chi\bigg(\frac{w_1}{\epsilon_1} \bigg)
\chi\bigg(\frac{w_2}{\epsilon_2} \bigg) e^{-(v,\,w) \Sigma^{-1}
(v,\, w)^T/2}\frac{dvdw}{\sqrt{\det{\Sigma}}(2\pi)^{d+1}} \\&=
\frac{1}{4\epsilon_1\epsilon_2}
\int_{-\epsilon_1}^{\epsilon_1}\int_{-\epsilon_2}^{\epsilon_2}
\int_{\R^{2d}} \|v_1\|
\|v_2\| e^{-(v,\,w) \Sigma^{-1} (v,\,
w)^T/2}\frac{dvdw}{\sqrt{\det{\Sigma}}(2\pi)^{d+1}} \;.
\end{split}
\end{equation}

%Note that the values of $K(x,y)$ on a set of measure
%zero do not have any impact on the value of the integrals
%involved, and we will neglect it from this point on.

We therefore have
\begin{multline*}
K(x,y) = \\
\lim_{\epsilon_1, \epsilon_2\to 0}
\frac{1}{4\epsilon_1\epsilon_2}
\int_{-\epsilon_1}^{\epsilon_1}
\int_{-\epsilon_2}^{\epsilon_2} \int_{\R^{2d}} \|v_1\|
\|v_2\| e^{-(v,\,w) \Sigma^{-1} (v,\,
w)^T/2}\frac{dvdw}{\sqrt{\det{\Sigma}}(2\pi)^{d+1}} \;.
\end{multline*}
Since the last integrand is continuous, we may use the
fundamental theorem of the calculus to replace the averaging
over $w_1$, $w_2$ by the value at $w_1=w_2=0$, to obtain
$$
K(x,y) = \int_{\R^{2d}} \|v_1\|\|v_2\|
e^{-(v,\,\vec 0) \Sigma^{-1} (v,\, \vec 0)^T/2}
\frac{dvdw}{\sqrt{\det{\Sigma}}(2\pi)^{d+1}} \;.
$$

We have
$$
\Sigma^{-1} = \begin{pmatrix}
          *&*\\ * &\Omega^{-1}
          \end{pmatrix}
$$
with $\Omega= C-B^TA^{-1}B$ being the reduced covariance matrix, which
we computed in \eqref{expression for Omega}. Together with
\eqref{eq:Jacobi's formula}  we find
\begin{equation*}
\begin{split} K(x,y) &=  \frac 1 {\sqrt{1-u^2}} \int_{\R^{2d}}\
| v_1| | v_2| \frac {\exp(-\frac 12  v\Omega^{-1}  v^T)}
{\sqrt{\det\Omega}} \frac{d v}{(2\pi)^{d+1}} \;.
\end{split}
\end{equation*}

Finally, we get Proposition~\ref{prop:justif ord chng}
by noticing that $K(x,y)=K(x-y)$,
%(that is, "the random process is stationary"),
and therefore the (double)
integral in \eqref{eq:int form sec mom dbl} may be expressed as a
single integral.
\end{proof}

In the course of the proof we saw that
$
K(x,y)=\lim_{\epsilon_1,\epsilon_2\to 0}K_{\epsilon_1,\epsilon_2}(x)\;.
$
Therefore, taking the limit $\epsilon_1,\epsilon_2\to 0$ and using
Lemma~\ref{lem:bound on K eps(x,y)}   we obtain
\begin{corollary}\label{cor:ker bnd ler ker}
If $u(x)^2 \ne 1$ then
\begin{equation*}
K(x) \ll \frac{\eigenvalue}{\sqrt{1-u(x)^2}}\;.
\end{equation*}
\end{corollary}

\section{A bound for the variance}\label{sec:variance}
In this section we prove:
\begin{theorem}\label{thm:var bnd}
For $d\geq 2$, \begin{equation*} \var(Z) =
O\bigg( \frac{\eigenvalue}{\sqrt{\Ndim}}\bigg).
\end{equation*}
\end{theorem}
\subsection{Plan of the proof}
We use the integral formula \eqref{eq:int form sec mom} for the
second moment of $Z(f)$, that is
$\E(Z^2)  = \int_{\TT^d} K(z)dz$,
with
$$
K(z)=\frac 1 {\sqrt{1-u^2}} \int_{\R^{2d}}\ \| v_1\| \| v_2\|
\frac {\exp(-\frac 12  v\Omega^{-1}  v^T)} {\sqrt{\det\Omega}}
\frac{d v}{(2\pi)^{d+1}}\;.
$$
As in \cite{ORW}, we will define a notion of ``singular
points'' in $\TT^d$ where the factor $1/\sqrt{1-u^2}$
is large, and treat separately the singular and
nonsingular points. The singular set is shown to give
a contribution of $O(\eigenvalue/\Ndim)$.
% in fact except in dimension $d=3,4$ it gives $o(\eigenvalue/\Ndim)$.
On the nonsingular set, the factor $1/\sqrt{1-u^2}$ may, up to an
admissible error,  safely be replaced by $1$. To treat
the Gaussian integral, we write
$$\Omega(z)=\frac{4\pi^2\eigenvalue}{d} (I-S(z))$$
and recover the square of the expected value $\E(Z)^2$ from the
contribution of the identity matrix $I$;  the rest is then the key
quantity for bounding the variance. Setting $\sigma(z)$ to be the
spectral norm  of $S(z)$, we show that that variance is bounded by
$\eigenvalue\left(\int_{\TT^d}\sigma(z)dz + O(1/\Ndim)\right)$.
Now $\sigma(z)$ is at most  $\sqrt{\tr(S(z)^2)}$, whose integral
we need to bound. We do this by using Cauchy-Schwartz, which
allows us to bound it  by $(\int_{\TT^d} \tr(S(z)^2) dz)^{1/2}\ll
1/\sqrt{\Ndim}$. Hence the variance is $\var(Z)\ll
\eigenvalue/\sqrt{\Ndim}$. It should be possible to improve this
to $O(\eigenvalue/\Ndim)$.
%\begin{verbatim}
%do we believe this ?
%\end{verbatim}

\subsection{The singular set}\label{subsec:singular}
We give the definition of \cite{ORW} for singular points:
\begin{defn}
A point $x\in \T^d$ is a positive singular point if there
is a set of frequencies $\Lambda_x\subset \Lambda$ with density
$\frac{|\Lambda_x|}{|\Lambda|}>1-\frac 1{4d}$ for which $\cos 2\pi
\langle \lambda, x \rangle >3/4$ for all $\lambda\in \Lambda_x$.
Similarly we define a negative singular point to be a point $x$
where there is a set $\tilde\Lambda_x\subset \Lambda$ of density
$>1-\frac 1{4d}$ for which $\cos 2\pi\langle \lambda, x\rangle <-3/4$ for
all $\lambda\in \tilde\Lambda_x$.
\end{defn}

Let $M\approx \sqrt{\eigenvalue}$ be a large integer.
%\footnote{It suffices to take $M=\lfloor 16\pi \sqrt{d}\sqrt{E} \rfloor$.}.
We decompose the torus $\TT^d=\R^d/\Z^d$ as a disjoint union (with
boundary overlaps) of $M^d$ closed cubes $I_{\vec k}$ of side length
$1/M$ centered at $\vec k/M$, $\vec k\in \Z^d$.

\begin{defn}A cube $I_{\vec k}$ is a positive (resp. negative)
singular cube if it contains a positive (resp. negative) singular
point.
\end{defn}

\begin{defn}
The singular set $\sing$ is the union of all singular cubes.
\end{defn}

In \cite{ORW}, we showed that the measure of the singular set is
bounded by
\begin{equation}\label{eq: bounding meas B}
\meas(\sing) \ll \int_{\TT^d} u(x)^4 dx \ll \frac 1{\Ndim}
\end{equation}
(and except in dimensions $d=3,4$ this is $o(1/\Ndim)$).

In order to bound the contribution of the singular set to the
integral in \eqref{eq:int form sec mom}, we use
Corollary~\ref{cor:ker bnd ler ker}. It was shown in \cite{ORW}
(see (6.3)) that
\begin{equation}
\int_{\sing} \frac{dx}{\sqrt{1-u(x)^2}} \ll
\int_{\T^d} u(x)^4dx \ll\ \frac 1{\Ndim}\;.
\end{equation}
Therefore we obtain:
\begin{corollary}\label{cor:sing set}
The contribution of the singular set is bounded by
$$
\int\limits_{\sing} K(x) dx \ll \eigenvalue\int_{\T^d} u(x)^4 dx
\ll\frac{\eigenvalue}{\Ndim} \;.
$$
\end{corollary}

\subsection{The nonsingular set}
We now want to estimate the contribution of the nonsingular set to the
integral formula of Proposition~\ref{prop:justif ord chng}
for the second moment of $Z$. Recall that it reads $\E(Z^2) =
\int_{\TT^d} K(z)dz$ with the kernel $K(z)$ given by
\eqref{eq:formula for K(x)}, that is
$$K(z)= \frac 1 {\sqrt{1-u(z)^2}} \int_{\R^{2d}}\ \| v_1\| \| v_2\|
\frac {\exp(-\frac 12  v\Omega(z)^{-1}  v^T)}
{\sqrt{\det\Omega(z)}} \frac{d v}{(2\pi)^{d+1}}\;.
$$

A consequence of the definition of singular points is that on the
nonsingular set, $u$ is bounded away from $\pm 1$.
In  \cite[lemma 6.5]{ORW}  we showed that if $x\in \TT^d$ is nonsingular then
\begin{equation*}
\label{lem:bounding u}
|u(x)|<1-\frac 1{16d} \;.
\end{equation*}
As a consequence, on the nonsingular set, we may expand
$$
\frac 1{\sqrt{1-u^2}} = 1+O(u^2) \;,
$$
where the implied constant depends only on $d$.

We now wish to handle the ``reduced covariance matrix''
$\Omega$ of \eqref{expression for Omega} on the nonsingular set.
We write $\Omega=(4\pi^2\eigenvalue/d)\cdot \Omega_1$ and
$\Omega_1= I-S$, where
\begin{equation}\label{eq:formula for S}
S= \frac {d}{4\pi^2 \eigenvalue} \frac 1{1-u^2}
\begin{pmatrix}  D^T D& (1-u^2)H + uD^TD\\(1-u^2)H +uD^T D& D^TD
\end{pmatrix}
\end{equation}
Note that since outside a set of measure zero, $\Omega_1\gg 0$ is
positive definite,
we have $S\ll I$ in the sense that all eigenvalues of $S$ are
in $(-\infty,1)$. Let $\sigma$ be the {\em spectral norm} of $S$, so
that denoting the eigenvalues of $S$ by
$\alpha_1,\dots,\alpha_{2d}$,
$$\sigma = \max_{1\leq j\leq 2d} |\alpha_j| \;.
$$
We give a bound on the mean and the mean-square of $\sigma$ on the
complement $B^c$ of the singular set.
\begin{lemma}\label{lem: bound for sigma, sigma^2}
\begin{equation}\label{eq: bd for sigma^2}
\int_{\nonsing} \sigma^2 dx \ll \frac 1{\Ndim}
\end{equation}
and
\begin{equation}\label{eq: bd for sigma}
\int_{\nonsing} \sigma dx \ll \frac 1{\sqrt{\Ndim}} \;.
\end{equation}
\end{lemma}
\begin{proof}
The bound \eqref{eq: bd for sigma}  follows from \eqref{eq: bd for
sigma^2} by applying Cauchy-Schwartz, so it suffices to prove
\eqref{eq: bd for sigma^2}. We have
$\sigma^2\leq \sum \alpha_j^2 = \tr(S^2)$,
and so it suffices to show
\begin{equation}
\int_{\nonsing} \tr(S^2) dx \ll \frac 1{\Ndim} \;.
\end{equation}

On the nonsingular set, the expression $\frac 1{1-u^2}$ is bounded,
and hence for purposes of upper bounds may be ignored. The entries
of $S^2$ on the nonsingular set are thus bounded by sums of the
following expressions :
$$
\frac 1{\eigenvalue^2} \frac{\partial^2 u}{\partial x_i \partial
x_j} \frac{\partial^2 u}{\partial x_k \partial x_\ell}, \quad \frac
1{\eigenvalue^2} \frac{\partial^2 u}{\partial x_i \partial
x_j}\frac{\partial u}{\partial x_k} \frac{\partial u}{\partial
x_\ell},\quad \frac{1}{\eigenvalue^2} \frac{\partial u}{\partial
x_i} \frac{\partial u}{\partial x_j} \frac{\partial u}{\partial x_k}
\frac{\partial u}{\partial x_\ell}
$$
and it suffices to show that the integral of each over all of
$\TT^d$ is $O(1/\Ndim)$.

By applying Cauchy-Schwartz, it suffices to show
$$
\int_{\TT^d} (\frac{\partial^2 u}{\partial x_i \partial x_j})^2 dx
\ll \frac {\eigenvalue^2}{\Ndim}
\qquad \mbox{and} \qquad
\int_{\TT^d} (\frac{\partial u}{\partial x_k} )^4 dx
\ll \frac {\eigenvalue^2}{\Ndim} \;.
$$

We have
$$\frac{\partial^2 u}{\partial x_i \partial x_j} =
\frac{-8\pi^2}{\Ndim}\sum_{\lambda \in \Lambda/\pm} \lambda_j
\lambda_k \cos 2\pi \langle \lambda, x\rangle
$$
and hence
\begin{equation*}
\begin{split}
\int_{\TT^d} (\frac{\partial^2 u}{\partial x_i \partial x_j})^2 dx
  &= (\frac{8\pi^2}{\Ndim})^2 \sum_{\lambda, \mu \in
\Lambda/\pm} \lambda_i \lambda_j \mu_i \mu_j
\frac {1}{2}\delta(\lambda,\mu)\\
&\ll \frac 1{\Ndim^2} \sum_{\lambda\in \Lambda} \lambda_i^2
\lambda_j^2 \ll \frac{\eigenvalue^2}{\Ndim}
\end{split}
\end{equation*}
since $\lambda_j^2\leq |\lambda |^2 = \eigenvalue$.

To bound  $\int_{\TT^d} (\frac{\partial r}{\partial x_k})^4 dx$, we write
$$\frac{\partial u}{\partial x_k} =
\frac{-4\pi}{\Ndim}\sum_{\lambda \in \Lambda/\pm} \lambda_k
\sin  2\pi \langle \lambda, x\rangle \;,
$$
and as above, we have
$$
\int_{\TT^d} (\frac{\partial u}{\partial x_k})^4 dx \ll
\frac{1}{\Ndim^4}
\sum\limits_{\substack{\lambda^1,\ldots,\lambda^4\in\Lambda/\pm
\\ \lambda^1\pm\lambda^2\pm\lambda^3\pm\lambda^4=0}} \lambda^1_k\cdot
\lambda^2_k\cdot \lambda^3_k\cdot \lambda^4_k \ll
\frac{\eigenvalue^2}{\Ndim}\;,
$$
since $\lambda^1,\lambda^2,\lambda^3$ determine $\lambda^4$ once
we decree that $\lambda^1\pm\lambda^2\pm\lambda^3\pm\lambda^4=0$,
and $|\lambda^i_k| \ll \sqrt{\eigenvalue}$.
%As is explained in \cite{ORW}, except perhaps in dimensions
%$d=3,4$, we may reduce this to $o(\eigenvalue/\Ndim)$.
\end{proof}

\subsection{Concluding the proof of Theorem~\ref{thm:var bnd}}
Since $\Omega_1=\Omega \cdot d/(4\pi^2\eigenvalue)$ is symmetric and
positive definite (away from a set of measure zero), it has a positive definite
square root $P_1=P_1^T\gg 0$, $\Omega_1 = P_1^2$.
By proposition ~\ref{prop:justif ord chng}
\begin{equation*}
\begin{split}
K(x) &=  \frac 1 {\sqrt{1-u^2}} \int_{\R^{2d}}\ |\vec v_1| |\vec
v_2| \frac {\exp(-\frac 12 \vec v\Omega^{-1} \vec v^T)}
{\sqrt{\det\Omega}} \frac{d\vec v}{(2\pi)^{d+1}} \\
&=\frac{4\pi^2\eigenvalue}{d}  \frac {1} {\sqrt{1-u^2}}
\int\limits_{\R^{2d}} |\vec (z P_1)_1| |(\vec z P_1)_2| e^{-|\vec
z|^2/2} \frac{d\vec {z}}{(2\pi)^{d+1}}
\end{split}
\end{equation*}
on using the change of variables
$\vec v = \frac{2\pi\sqrt{\eigenvalue}}{\sqrt{d}} \cdot \vec {z} \cdot
P_1 $.

We claim that
$$P_1 = I\left( 1 +O(\sigma) \right) \;.
$$
Indeed,  if $S=UDU^T$, $U$ orthogonal and
$D=\mbox{diag}(\alpha_1,\dots,\alpha_{2d})$ then $P_1 =
U(I-D)^{1/2}U^T$ and using the inequality
$|\sqrt{1-\alpha}-1|<|\alpha|$ for $-\infty<\alpha<1$, gives
$$ (I-D)^{1/2}= I
+O(\begin{pmatrix} |\alpha_1|& &\\ &\ddots&\\& &|\alpha_{2d}|
\end{pmatrix})  = I(1+O(\sigma)) \;.
$$
Thus we may write $zP_1 = z\left(1+O(\sigma)\right)$.

On the nonsingular set, we may expand
$$\frac 1{\sqrt{1-u^2}} = 1+O(u^2)$$
and so we find that on the nonsingular set %(recall $|u|\leq 1$)
\begin{equation*}
\begin{split}
K(x) &= \frac{4\pi^2\eigenvalue}{d}\int_{\R^{2d}} |\vec z_1| |\vec
z_2| e^{-|\vec z|^2/2} \left( 1+O(u^2)\right)
\left(1+O(\sigma)  \right)^2 \frac{d\vec {z}}{(2\pi)^{d+1}} \\
&= \E(Z)^2 \left ( 1+O(u^2) +O(\sigma)+O(\sigma^2) \right) \;.
\end{split}
\end{equation*}
Integrating over the nonsingular set,  and using
$$\int_{\nonsing} 1 = 1+O(\meas(\sing))$$
%bounded to replace the nonsingular set with all of $\TT^d$,
we find
\begin{equation*}
\frac {\int_{\nonsing} K(x)dx}{\E(Z^2)} = 1+
O\bigg(\int_{\nonsing}(u^2 +\sigma +\sigma^2) dx \bigg) +
O(\meas(\sing)) \;.
\end{equation*}

Now
$
\int_{\TT^d}u(x)^2 dx =\frac 1{\Ndim} \;,
$
and by lemma~\ref{lem: bound for sigma, sigma^2}
$$
\int_{\nonsing}\sigma^2 dx\ll \frac 1{\Ndim}\;, \qquad
\int_{\nonsing}\sigma dx  \ll \frac 1{\sqrt{\Ndim}}\;.
$$
Furthermore, by \eqref{eq: bounding meas B},
$$
\meas(\sing) \ll \int_{\TT^d} u^4 dx \ll \frac 1{\Ndim} \;,
$$
so that we find
$$
\int_{\nonsing} K(x) dx  = \E(Z^2) \left(1 +O(\frac1{\sqrt{\Ndim}})
\right) \;.
$$
By Corollary ~\ref{cor:sing set}, the singular set contributes at
most
$$
\int_{\sing} K(x)dx \ll \frac{\eigenvalue}{\Ndim}\;.
$$
%(and even better except in dimensions $d=3,4$).
Therefore we find
$$
\E(Z^2) = \E(Z)^2 +O(\frac{\eigenvalue}{\sqrt{\Ndim}})\;,
$$
that is
$$
\var(Z) \ll \frac{\eigenvalue}{\sqrt{\Ndim}} \;.
$$
Thus we have concluded the proof of Theorem~\ref{thm:var bnd}. \qed

\appendix

\section{The non-degeneracy of the covariance matrix}
\label{sec:nondeg of cov}
In this appendix we show that the covariance matrix defined by
\eqref{eq:covariances of f} is nonsingular for almost all
$(x,y)\in (\T^d)^2$, thereby justifying the change of variables
\eqref{eq: change vars}.

\begin{proposition}\label{prop:gen nondeg proc}
Assume that $\Ndim \gg_d 1$ and $d\geq 2$.
Then for almost all $(x,y)\in \T^d\times \T^d$ the
linear map $\eigenspace\rightarrow \R^{2d+2}$ defined by
$$f\mapsto\big(f(x),\, f(y),\, \nabla f(x),\, \nabla f(y)\big),$$
is surjective.
\end{proposition}

We want to show that for  almost all pairs $(x,y)\in \TT^d\times \TT^d$, the
only  vector $(\alpha,\beta,\vec C,\vec D)\in \R^{2d+2}$ satisfying
$$
\alpha f(x) +\beta f(y) +\frac 1{2\pi}\langle C, \nabla f(x) \rangle +
\frac 1{2\pi}\langle D, \nabla f(y) \rangle =0, \qquad \forall f\in
\eigenspace
$$
is the zero vector. Taking $f(x) = e^{2\pi i \langle \lambda,x
\rangle}$, $\lambda\in \Lambda$  gives
$$
\alpha e^{2\pi i \langle \lambda,x\rangle} +
\beta e^{2\pi i \langle\lambda,y\rangle} +
i e^{2\pi i \langle \lambda,x\rangle} \langle C, \lambda \rangle +
i e^{2\pi i \langle \lambda,y\rangle} \langle D, \lambda \rangle =
0,\qquad \forall \lambda \in \Lambda
$$
or setting $z=y-x$,
$$
\alpha +i \langle C,\lambda \rangle  = -e^{2\pi i \langle
\lambda,z\rangle} \left( \beta+ i \langle D, \lambda \rangle
\right),\qquad \forall \lambda\in \Lambda \;.
$$
Thus we are reduced to proving the following:

\begin{lemma}
\label{lem:gen no sol}
Assume that $\Ndim \gg_d 1$ and $d\geq 2$.
Then for almost all $z$, the only solution for the
equation
\begin{equation}\label{eq:orig eq}
\alpha+i \langle C,\lambda \rangle  = - e^{2\pi
i\langle \lambda,z\rangle} \left(\beta+i \langle D,\lambda\rangle
\right) ,\qquad \forall \lambda\in \Lambda,
\end{equation}
is $\alpha=\beta=0$, $C=D=\vec 0$.
\end{lemma}

\begin{proof}
We divide the work into two steps: In the first step, we show that for
{\em all} $z\in \TT^d$,  the solutions of \eqref{eq:orig eq} satisfy
$\beta=\pm \alpha$ and $D=\pm C$. In the second step, we take
$\beta=\pm \alpha$ and $D=\pm C$  and show that for
{\em almost all} $z\in \TT^d$, the only solutions of
\eqref{eq:orig eq} are $\alpha=0$ and  $C=\vec 0$.

\bigskip
\noindent{\bf Step 1:} We first show that for all $z\in \TT^d$,
all solutions of \eqref{eq:orig eq} satisfy $\beta=\pm \alpha$ and
$D=\pm C$.

Taking  squared norms of both sides of \eqref{eq:orig eq}, we get
\begin{equation*}
\alpha^2+\langle C,\, \lambda \rangle^2 = \beta^2 +\langle D,\,
\lambda \rangle^2,\qquad \forall \lambda\in \Lambda
\end{equation*}
or
$$\langle C,\, \lambda \rangle^2 - \langle D,\,
\lambda \rangle^2 = \beta^2-\alpha^2, \qquad \forall \lambda\in
\Lambda \;.
$$
Setting $A = C-D=(a_1,a_2,\dots)$ and $B = C+D=(b_1,b_2,\dots)$, we have
\begin{equation}\label{eq:bas rel norms}
\langle A, \lambda\rangle \cdot \langle B, \lambda\rangle =
\beta^2-\alpha^2,\qquad \forall \lambda\in \Lambda
\end{equation}
and it suffices to see that $A=\vec 0$ or $B=\vec 0$.

If $\Ndim \gg_d 1$, then there is some $\lambda\in \Lambda$ with two
nonzero coordinates, say $\lambda_1\lambda_2\neq 0$ (by applying a
permutation of the coordinates to $\lambda$ we may replace $1$ and $2$ by any pair
of distinct indices).
For each $\epsilon\in \{\pm 1\}^d$, replace $\lambda$  in
\eqref{eq:bas rel norms}  by
$$
\lambda^\epsilon:=(\epsilon_1\lambda_1,\epsilon_2\lambda_2
,\dots, \epsilon_d \lambda_d) \;,
$$
multiply the result by
$$
\chi_{1,2}(\epsilon) =
\epsilon_1 \epsilon_2
$$
and sum the resulting equalities over all $\epsilon\in \{\pm 1\}^d$,
using
$$\sum_{\epsilon\in \{\pm 1\}^d} \chi_{1,2}(\epsilon)=0$$
to get
$$
\sum_{\epsilon\in \{\pm 1\}^d} \chi_{1,2}(\epsilon)
\langle A, \lambda^\epsilon\rangle \cdot \langle B, \lambda^\epsilon\rangle = 0\;.
$$
Expanding
$$
\langle A, \lambda^\epsilon\rangle \cdot \langle
B,\lambda^\epsilon\rangle = \sum_{j,k=1}^d\epsilon_j\epsilon_k a_j b_k\lambda_j\lambda_k
$$
and using
$$
\sum_{\epsilon\in \{\pm 1\}^d}\chi_{1,2}(\epsilon) \epsilon_j\epsilon_k =
\begin{cases}
2^d,& (j,k)=(1,2) \mbox{ or } (2,1) \\ 0& \mbox{ otherwise }
\end{cases}
$$
we get
$$ 2^d \lambda_1 \lambda_2 (a_1b_2+a_2b_1) = 0$$
and since we assume $\lambda_1\lambda_2\neq 0$, we find
$a_1b_2+a_1b_2=0$. Repeating the argument with any pair of distinct
indices finally shows that
\begin{equation}\label{eq:ij}
a_ib_j+a_jb_i = 0,\quad \forall i\neq j\;.
\end{equation}

If $A\neq \vec 0$, say $a_1\neq 0$, then we find that
\begin{equation}\label{eq:bj for aj}
b_j = -\frac{b_1}{a_1}a_j ,\qquad \forall j\neq 1 \;.
\end{equation}
Thus if $b_1=0$ then all $b_j=0$, that is $B=\vec 0$ and we are done.
Therefore we may assume that $b_1\neq 0$ (and we have also assumed
$a_1\neq 0$). We will show this cannot happen.

If $d>2$, we substitute \eqref{eq:bj for aj} in \eqref{eq:ij}
with  any $i\neq 1$, $j\neq 1$ to get
$$
2\frac{b_1}{a_1} a_ia_j = 0
$$ that is since $b_1\neq 0$, that
$$
a_ia_j=0,\qquad \forall i\neq j,\quad i,j\neq 1 \;.
$$
Thus there is at most one index $k\neq 1$ with $a_k\neq 0$, say $k=2$,
so we find that $a_j=0$ for $j\neq 1,2$, and by \eqref{eq:bj for aj}
we therefore have $b_j=0$  for $j\neq 1,2$. Thus
$$A=(a_1,a_2,\vec 0),\quad B= \frac{b_1}{a_1} (a_1,-a_2,\vec 0)$$
(if $d=2$ this still holds, we just ignore the extra coordinates).

Plugging this into \eqref{eq:bas rel norms}  with $\lambda$ so that
$\lambda_1\neq \pm \lambda_2$ (which exists if $\Ndim \gg_d 1$) gives
\begin{equation}\label{eq:diff 1}
(a_1 \lambda_1)^2-(a_2 \lambda_2)^2 = \frac{a_1}{b_1}(\beta^2-\alpha^2)
\end{equation}
and replacing $\lambda=(\lambda_1,\lambda_2,\dots)$ with
$(\lambda_2,\lambda_1,\dots)$ gives
\begin{equation}\label{eq:diff 2}
(a_1 \lambda_2)^2-(a_2 \lambda_1)^2 =
\frac{a_1}{b_1}(\beta^2-\alpha^2) \;.
\end{equation}
Comparing \eqref{eq:diff 1} with \eqref{eq:diff 2} gives
$$(a_1 \lambda_1)^2-(a_2 \lambda_2)^2 = (a_1 \lambda_2)^2-(a_2
\lambda_1)^2
$$
that is
$$ (\lambda_1^2-\lambda_2^2)(a_1^2+ a_2^2) = 0 \;.$$
Since we chose $\lambda_2\neq \pm \lambda_1$ this gives $a_1=a_2=0$,
contradicting $a_1\neq 0$. Thus we are done with step 1.

\bigskip
\noindent{\bf Step 2:}
We take  $C=\pm D$ and
$\alpha=\pm\beta$ in \eqref{eq:orig eq} and wish to show that for
almost all $z\in \TT^d$, the only solutions are $\alpha=0$ and $C=\vec
0$.
If either ($\alpha = \beta$ and $C = D$) or
($\alpha = -\beta$ and $C=-D$), then \eqref{eq:orig eq} gives
$e^{2\pi i \langle z,\lambda\rangle}=-1$ which is a measure zero condition.

Otherwise, assume $\alpha=\beta$ and $C=-D$ (the other case is treated
similarly). Here we have
\begin{equation}
\label{eq:bas rel C=D}
 \alpha+i \langle C,\lambda \rangle  = -
e^{2\pi i\langle \lambda,z\rangle} \left(\alpha-i \langle
C,\lambda\rangle \right).
\end{equation}
If $\alpha=0$ and $C\neq 0$ then there is some $\lambda\in \Lambda$
so that $\langle C,\lambda \rangle\neq 0$ and \eqref{eq:bas rel C=D}
forces $e^{2\pi i \langle z,\lambda\rangle}=1$, that is $z$ lies on
one of the hyperplanes $$\cup_{\lambda\in \Lambda}\{z:\langle
\lambda,z\rangle=0\mod 1 \},$$ which is a measure zero condition.

If $\alpha\neq 0$, we replace  $C$ by $-\frac{1}{\alpha} C$ and
$\alpha$ by $1$ and drop the negative sign.
Taking  the real part of \eqref{eq:bas rel C=D}, we have
$$1+\cos 2\pi \langle \lambda, z \rangle =
\langle C, \lambda\rangle \sin 2\pi \langle \lambda, z \rangle \;.
$$
We may assume that the sine on
the RHS doesn't vanish, since $\sin 2\pi \langle \lambda, z
\rangle=0$ is a measure zero condition. Therefore, we may divide to
get
$$
\langle C,\, \lambda \rangle  =
\frac{1+\cos 2\pi \langle\lambda, z \rangle}{\sin 2\pi \langle \lambda, z \rangle} =
 \cot \pi \langle \lambda, z \rangle\;. $$

Now square and average the result over an orbit $\mathcal O\subset
\Lambda$  of the group $W_d$ of all permutations and sign changes of
the coordinates. The LHS gives
$$
\frac 1{|\mathcal O|}  \sum_{\lambda\in \mathcal O} \langle
C,\lambda\rangle^2  = \frac{\eigenvalue}{d} | C|^2
$$
by \eqref{ORW, Lemma 5.2},
which is independent of the orbit chosen. The RHS gives
$$
\frac 1{|\mathcal O|}  \sum_{\lambda\in \mathcal O} \cot^2\pi
\langle \lambda,z\rangle = 1+\frac 1{\mathcal O} \sum_{\lambda\in
\mathcal O} \frac 1{(\sin \pi \langle \lambda,z\rangle)^2}
$$
that is we find
\begin{equation*}
\frac{\eigenvalue}{d} | C|^2-1 = \frac 1{|\mathcal O|}
\sum_{\lambda\in \mathcal O} \frac 1{(\sin \pi \langle
\lambda,z\rangle)^2}\;.
\end{equation*}
Since $1/(\sin \pi \langle \lambda,z\rangle)^2$ is even, we get the
same term for $\lambda$ and $-\lambda$ and so we may  replace the
average over $\mathcal O$ by the average over $\mathcal O/\pm$ where
we have taken only one of $\lambda,-\lambda$. Thus
\begin{equation}\label{eq:single orbit}
\frac{\eigenvalue}{d} | C|^2-1 = \frac 1{|\mathcal O/\pm|}
\sum_{\lambda\in \mathcal O/\pm} \frac 1{(\sin \pi \langle
\lambda,z\rangle)^2}\;.
\end{equation}

Assuming that $\Ndim> |W_d|=2^d d!$, we can find a {\em different}
orbit $\mathcal O'\subset \Lambda$ and then comparing with
\eqref{eq:single orbit} gives
\begin{equation}\label{eq:dependence}
\frac 1{|\mathcal O/\pm|} \sum_{\lambda\in \mathcal O/\pm} \frac
1{(\sin \pi \langle \lambda,z\rangle)^2} = \frac 1{|\mathcal
O'/\pm|} \sum_{\lambda\in \mathcal O'/\pm} \frac 1{(\sin \pi \langle
\lambda,z\rangle)^2}
\end{equation}
that is we have eliminated the variable $C$.

We claim that \eqref{eq:dependence} forces the point $z$ to lie on a
measure zero subset of $\TT^d$. Indeed, the functions involved are
meromorphic in $\C^d/\Z^d$ and hence if \eqref{eq:dependence} does
not hold for all $z$, it can only hold on a complex submanifold of
codimension (at least) one and in particular its real points will
have codimension at least one in $\TT^d$. But near the origin $z=0$,
each of the functions $1/(\sin \pi \langle \lambda,z\rangle)^2$ has
singularities on the hyperplane $\langle z,\lambda \rangle=0$ and
these hyperplanes are distinct for  $\lambda$'s which are not
collinear (here the condition $d\geq 2$ comes in), as is the case
for those appearing in \eqref{eq:dependence}. Thus these functions
are linearly independent and so \eqref{eq:dependence} is not valid
for all $z$.
\end{proof}

\end{document}